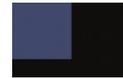

# Bow-Tie Cavity for Terahertz Radiation

**Luigi Consolino** [1,*], **Annamaria Campa** [1], **Davide Mazzotti** [1], **Miriam Serena Vitiello** [2], **Paolo De Natale** [1] **and Saverio Bartalini** [1]

1   CNR–Istituto Nazionale di Ottica and LENS (European Laboratory for Non-linear Spectroscopy), Via N. Carrara 1, 50019 Sesto Fiorentino (FI), Italy; annamaria.campa@ino.it (A.C.); davide.mazzotti@ino.it (D.M.); paolo.denatale@ino.it (P.D.N.); saverio.bartalini@ino.it (S.B.)
2   NEST, CNR–Istituto Nanoscienze and Scuola Normale Superiore, Piazza San Silvestro 12, 56127 Pisa, Italy; miriam.vitiello@sns.it
*   Correspondence: luigi.consolino@ino.it; Tel.: +39-055-457-2292

**Abstract:** We report on the development, testing, and performance analysis of a bow-tie resonant cavity for terahertz (THz) radiation, injected with a continuous-wave 2.55 THz quantum cascade laser. The bow-tie cavity employs a wire-grid polarizer as input/output coupler and a pair of copper spherical mirrors coated with an unprotected 500 nm thick gold layer. The improvements with respect to previous setups have led to a measured finesse value $F$ = 123, and a quality factor $Q$ = 5.1·$10^5$. The resonator performances and the relevant parameters are theoretically predicted and discussed, and a comparison among simulated and experimental spectra is given.

## 1. Introduction

Since their invention, optical resonators have represented an important optical tool for their role as a key element for laser action, for the investigation of coherent radiation properties, and for all purposes that require control and enhancement of laser light. For almost all the electromagnetic spectrum, the physics of resonators has been well investigated with the experimental realization of very-high-quality-factor/finesse cavities, disclosing a large number of possible configurations [1–9]. However, among all possible applications, cavity-enhanced spectroscopy has found increasing popularity [10]. In fact, thanks to the capability of confining and enhancing optical power inside a cell, and thanks to the development of advanced techniques, such as cavity ring-down spectroscopy (CRDS) or saturated-absorption CRDS, high-finesse resonators have been able to reach sensitivity values down to parts per quadrillion (ppq) in trace gas sensing [11,12].

In this regard, the possibility to develop resonant cavities for terahertz (THz) light has long been sought. High-sensitivity and high-accuracy spectroscopy of rotational and ro-vibrational molecular transitions is indeed a key application in the terahertz (THz) spectral region, where transition line intensities can be very high, even with respect to those in the IR and microwave regions. Nevertheless, if compared to other spectral regions, achievements of THz spectroscopy are still quite scarce, due to the lack of proper tools, such as high-power, tunable lasers, fast, sensitive, and reliable detectors, and also of a set of performing optical elements, such as optical isolators, highly reflective mirrors or coatings and so on. A number of experimental setups, based on different approaches, have been recently developed in order to push forward the limits of THz spectroscopy [13,14]. For example, in 2013, a Gunn-oscillator-based frequency multiplier chain provided 1 kHz accuracy in the determination of a molecular transition line center (relative accuracy $1·10^{-9}$) [15], but this setup could only be operated in the lower part of the THz spectrum. In 2014, frequency referenced THz quantum cascade lasers (QCLs) were successfully used for THz spectroscopy, achieving a relative accuracy of $10^{-9}$ [16], and virtually giving access to the frequency window up to 4.5 THz, but with a limited tunability given by the characteristic of the single device. More recently, a novel setup based on difference-frequency generation has been reported, showing $10^{-9}$ accuracy and a broad tunability from 1 to 7.5 THz [17,18], however, with the drawback of a low emitted power (~0.5 W). All these spectroscopic setups would hugely benefit from the development of high-performance THz resonators, that could help, in principle, to tear down the present-day limit of $10^{-9}$ relative accuracy.

Resonant cavities for terahertz frequency light have been only developed in the last few years, due to the challenging lack of suitable materials and optical components; the best results reported to date relies on a wire grid polarizer (WGP) as input-output coupler. When the polarization of the incoming light is parallel to the wires of a WGP, it acts as a mirror, but the small fraction of light leaking through the WGP allow coupling of the external radiation to the cavity. The first results were published in reference [19], achieving $Q$-factors as large as $10^5$ in the sub-THz range (around 300 GHz). In 2015, our group proposed WGP-based resonators providing $Q$-factors of the same order ($2.5 \cdot 10^5$), but at much higher frequencies, i.e., 2.55 THz [20]. In this work, two different cavity geometries were investigated, namely V-shaped and ring-shaped resonators. The V-shaped cavity was composed by a WGP and two spherical mirrors, which helped to minimize beam profile distortion. The main limitation of this approach was the unavoidable on-resonance optical feedback (OF) on the laser source, inherent in the cavity geometry. Conversely, the ring-shaped cavity approach, allowed for a zero-feedback resonance, but its performances were mainly limited by the presence of two 90° off-axis parabolic mirrors, leading to distortion of both shape and polarization of the cavity beam.

In this paper, we propose an alternative cavity setup that overcomes the limitations of previous geometries, achieving a higher finesse. The necessity to have a ring geometry that minimizes the presence of feedback, and the use of optical elements working at very small angles, such as plane or spherical mirrors, suggested the solution of a bow-tie configuration, which is, by the way, the most common configuration in other spectral regions.

Here we report on the experimental realization of a THz WGP-based bow-tie resonator, to which radiation from a 2.55 THz QCL is coupled. The devised cavity has been characterized, and its relevant parameters have been compared with the prediction of a theoretical model.

## 2. Materials and Methods

The bow-tie cavity (BTC), as its name suggests, is designed such that the light traverses a closed path, as shown in Figure 1. The light is injected through a WGP tilted at a small angle with respect to the direction of the beam propagation. A plane mirror (PM) is placed to reflect the beam towards a couple of identical spherical mirrors (SM), with the same effective focal length (200 mm), put in a position that closes the path of the beam on the WGP.

The angle of incidence $\theta$ of the light in cavity on the WGP, is defined by the geometry, i.e., by the distances between each reflective element ($l_i$, $i = 1, \ldots, 4$), yielding the following mathematical expression:

$$\theta = \frac{1}{2} arccos \left( \frac{l_1 + l_3}{2 l_2} \right)$$

which for our case ($l_1 = 85$ mm, $l_{2,4} = 130$ mm, $l_3 = 145$ mm) is equal to 13.9°, with a total cavity length of 490 mm.

It is straightforward that the use of small incidence angles minimizes the effect of aberrations and preserves the high reflectivity of the mirrors, which are designed and manufactured for normal incidence. The gold coating of all the mirrors of the setup is made by chemical deposition (electroplating), has a thickness of about 500 nm, and has no protective dielectric layer. The reflectivity of these mirrors is supposed to be close to 99.5% [21], but the measurements performed in the laboratory resulted in $R_M$ = (99.0 ± 0.4%). At the same time, measurements performed on the WGP resulted in the following value of reflectivity, transmission, and absorption: ($R_{WGP}$, $T_{WGP}$, $A_{WGP}$) = (99.2%, 0.4%, 0.4%) [5].

The bow-tie cavity (BTC), as its name suggests, is designed such that the light traverses a closed path, as shown in Figure 1. The light is injected through a WGP tilted at a small angle with respect to the direction of the beam propagation. A plane mirror (PM) is placed to reflect the beam towards a couple of identical spherical mirrors (SM), with the same effective focal length (200 mm), put in a position that closes the path of the beam on the WGP.

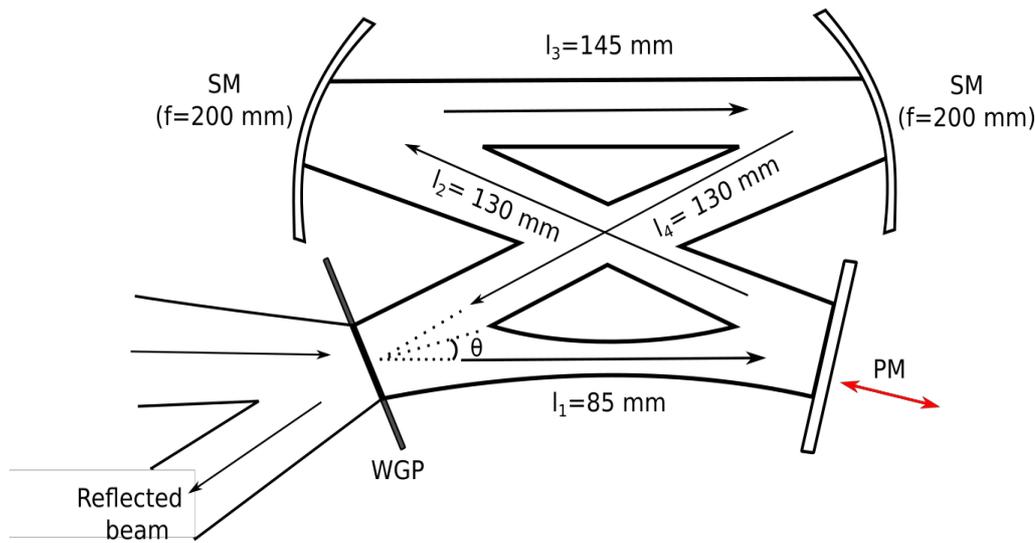

**Figure 1.** Configuration of the bow-tie cavity consisting in a plane gold mirror, two gold spherical mirrors (SM), and a wire grid polarizer (WGP) as input/output coupler. The red arrows indicate the motorized translating stage for the plane gold mirror.

The absorption in air at 2.55 THz due to water vapor should be included in the total losses of the BTC, but it can be strongly reduced by purging the cavity with nitrogen. In absence of absorption from air and by taking into account the physical properties of the WGP and of the mirrors, we can define a theoretical finesse for the BTC as [22,23]:

$$F = q\frac{\pi}{1 - R_{WGP} R_M^3}$$

returning a value of about 166.

The estimation of the main BTC parameters has been made with the simulation program Finesse [24], which requires an exact description of the geometry and properties of each element of the cavity. The QCL beam can be described by a Gaussian beam ($TEM_{00}$) and transverse modes ($TEM_{nm}$) that contribute with a small fraction of power and enlarge the beam waist from 1.8 mm (for $TEM_{00}$) to 3.5 mm. The main theoretical parameters of the cavity are reported in Table 1, and are compared with the values retrieved from our previous geometries. At the same time Figure 2 shows a simulation of the BTC spectrum where the first higher-order transverse modes have been excited, besides the fundamental one. This general simulation will be used, in the following, for the analysis of the experimental BTC spectrum.

**Table 1.** Main paramen comparison with the ones for the V-shapedters calculated for the bow-tie cavity, i and ring cavities [20] (FSR: Free spectral range). All parameters are calculated in vacuum, apart from column $F_{air}$ reporting the finesse values taking into account the water vapor absorption coefficient of ~0.0038 cm$^{-1}$ [20].

| Cavity | *l* (mm) | FSR (MHz) | $F_{air}$ | F | $\Delta\nu$ (MHz) | Q-Factor |
|---|---|---|---|---|---|---|
| V-shape | 480 | 625 | 24 | 83 | 7.5 | $3.4 \cdot 10^5$ |
| Ring | 480 | 625 | 22 | 65 | 9.6 | $2.7 \cdot 10^5$ |
| BTC | 490 | 612 | 30 | 166 | 3.7 | $6.9 \cdot 10^5$ |

fundamental one. This general simulation will be used, in the following, for the analysis of the experimental BTC spectrum.

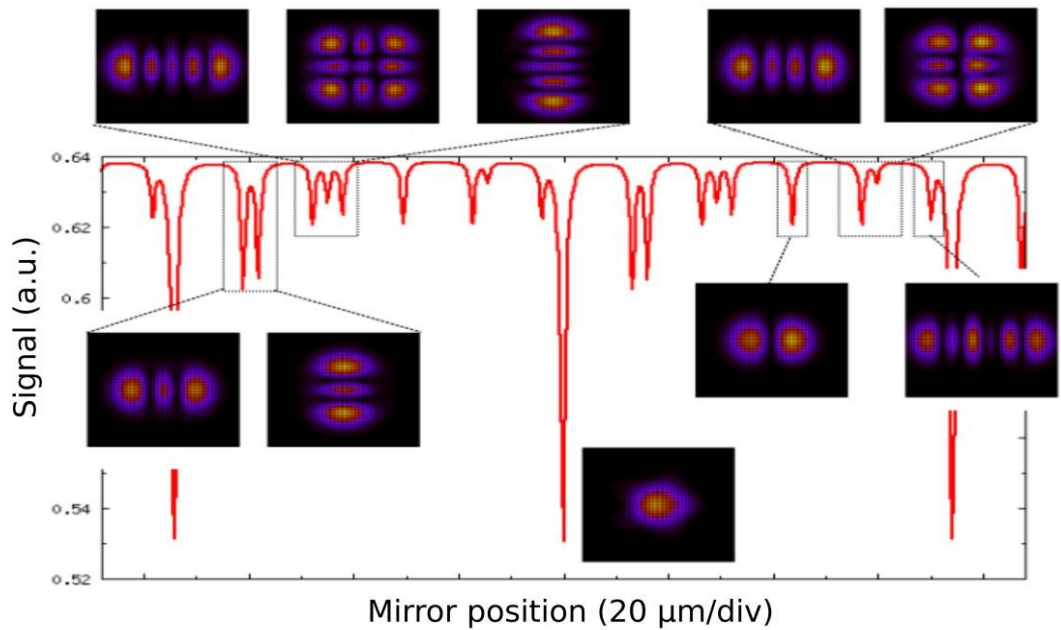

**Figure 2.** Simulation of the bow tie cavity (BTC) spectrum, showing contributions from higher-order transverse modes.

**Table 1.** Main parameters calculated for the bow-tie cavity, in comparison with the ones for the V-shaped and ring cavities [20] (FSR: Free spectral range). All parameters are calculated in vacuum, apart from column $F_{air}$ reporting the finesse values taking into account the water vapor absorption coefficient of ~0.0038 cm$^{-1}$ [20].

| Cavity / shape | FSR (MHz) | $F_m$ | FSR × $F$ (MHz) | $F_{air}$ | Q-factor |
|---|---|---|---|---|---|
| V-shape | 180 | 623 | 24 | 93 | 7.4 · $10^5$ |
| Ring | 480 | 625 | 22 | 65 | 9.6 · $10^5$ |
| BTC | 490 | 612 | 30 | 106 | 3.7 · $10^6$... 6.9 · $10^6$ |

## 3. Results

A sketch of the experimental setup is shown in Figure 3. The laser source is a QCL, with a bound-to-bound active region, emitting at 2.55 THz and mounted on the cold finger of a liquid helium cryostat. The QCL is driven in continuous-wave mode by a low-noise current driver at 370 mA with a fixed heat sink temperature T = 25 K. At this temperature the QCL threshold current is $I_{th}$ = 340 mA. The divergent QCL emission is collimated by an off-axis parabolic gold mirror, with an effective focal length of 25.4 mm, and guided by two plane mirrors to inject the cavity. To this purpose, the mode-matching conditions had to be satisfied by slightly focusing the beam towards the cavity with the parabolic mirror. Furthermore, the light emitted from a QCL can be described by an elliptical Gaussian beam with different divergences in the planes parallel and orthogonal to the epitaxial growth axis of its semiconductor heterostructure. Consequently, the mode-matching conditions had to be satisfied for both axes of the elliptical beam section. Between the collimating parabolic mirror and the cavity WGP, an optical system made of a half-wave plate (HWP) and a second wire-grid polarizer (WGP2) allows to choose the polarization of the electric field that will inject the cavity. In particular, the HWP allows us to rotate the linear QCL polarization, which is orthogonal to the growth axis, while the WGP2 cleans the polarization from residual components. The spectrum reflected by the cavity is then detected by a pyroelectric sensor aligned on the beam reflected by the WGP input coupler (QMC Instruments Ltd., Cardiff, UK, mod. QWG/RT), composed of tungsten wires with a diameter of 10 μm and spaced by 20 μm. The cavity is enclosed, by means of a plastic membrane, in nitrogen atmosphere, in order to purge out the water vapor that would induce additional intra-cavity losses.

In order to tune the cavity length, the plane mirror of the cavity is mounted on a motorized translation stage (Thorlabs Ltd., Ely, UK, mod. MTS25) controlled by a LabVIEW program, capable of setting the scanning speed (typical value 1.5 μm/s), the total scan length (typically 200 μm), the total scan time, and the acquisition rate. A chopper beam modulator and a lock-in amplifier allow us to acquire the reflected spectrum. In particular, when the cavity is off-resonance, the collected signal corresponds to the total incoming power on the WGP, while, when it is in resonance, a fraction of the light will be coupled to the cavity and a power dip is expected to appear in the signal. A different kind of acquisition can be performed by substituting the chopper modulation with a modulation of the translation stage (Thorlabs Ltd., Ely, UK, mod. MTS25) controlled by a LabVIEW program, capable of setting the scanning speed (typical value 1.5 µm/s), the total scan length (typically 200 µm), the total scan time, and the acquisition rate. A chopper beam modulator and a lock-in amplifier allow us to acquire the reflected spectrum. In particular, when the cavity is off-resonance, the collected signal corresponds to the total incoming power on the WGP, while, when it is in resonance, a fraction of the light will be coupled to the cavity and a power dip is expected to appear in the signal. A different kind of acquisition can be performed by substituting the chopper modulation with a modulation of the QCL driving current. In this case, when the BTC translation stage is tuned, the first derivative of the reflected spectrum is retrieved from the lock-in demodulation.

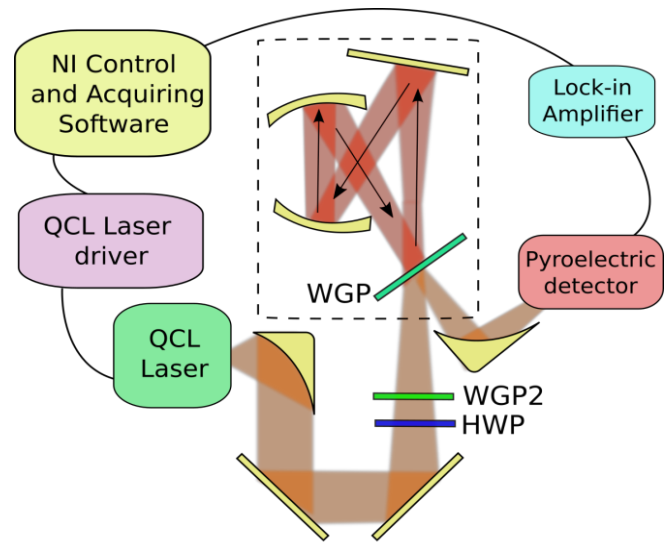

Figure 3. Experimental set-up. WGP and WGP2: Wire grid-polarizers, HWP: Half-wave plate.

## 4. Discussion

By measuring the beam waist at different distances from the cryostat it is possible to characterize the input beam parameters that can be regulated by finely moving the collimating parabolic mirror, in order to approach the mode-matching condition with the optical cavity. At that point, the alignment procedure is performed by looking at the experimental spectrum and comparing it with the simulated one reported in Figure 2. In this way, the transverse modes can be recognized, and the right corrections to the beam parameters can be applied, thus compensating either a wrong beam size or an off-axis alignment.

Moreover, as said before, in our setup it is possible to rotate the linear polarization of the field injected into the cavity. This allows making a systematic measurement of the width of the resonance peak for different orientations of the polarization, and therefore of the corresponding WGP wires direction, as shown in Figure 4. In this way it is possible to identify two different effects, appearing in the two orthogonal orientations. On one side, when the polarization is set to vertical, the presence of partial reflection from the cylindrical wires is expected to generate a disturbing optical feedback (OF) to the QCL. Indeed, we can experimentally observe the presence of this OF in the configuration with vertical polarization, as shown in Figure 4 (large angles) and in the right-inset. Here, a clear broadening of the resonance peak is present, due to the typical laser frequency pulling effect. The effect of OF is expected to disappear for horizontal orientation of the WGP wires, as shown in the left inset of Figure 4. However, in this configuration, there is an angle θ between the beam polarization and the horizontal orientation of the wires of the WGP. As a consequence, a fraction of the total field is not properly reflected, resulting in additional cavity losses, i.e., in a reduction of the finesse, as confirmed in Figure 4 (small angles). The trade-off between the two effects results in maximum measured finesse around an operation angle of 70°, in order to minimize OF on the laser, while being as close as possible to the vertical position, i.e., maximum WGP reflectivity.

In these conditions, i.e., with the nitrogen-purged air and with an orientation of the polarization at 70° inside the cavity, the BTC achieves a measured finesse of 123, corresponding to a Q-factor of $5.10^5$ and an enhancement factor of 11.6. An experimental acquisition in these conditions is shown in Figure 5. The retrieved finesse value is considerably lower than the theoretically expected one of 166, which could be measured at 90° polarization angle. This effect can be confirmed by comparing an experimental spectrum and a simulated one, shown in Figure 5 as well. The behavior of the acquired spectrum can In these conditions, i.e., with the nitrogen-purged air and with an orientation of the polarization at 70° inside the cavity, the BTC achieves a measured finesse of 123, corresponding to a Q-factor of 5.1·10[5] and an enhancement factor of 11.6. An experimental acquisition in these conditions is shown in Figure 5. The retrieved finesse value is considerably lower than the theoretically expected one of 166, which could be measured at 90° polarization angle. This effect can be confirmed by comparing an experimental spectrum and a simulated one, shown in Figure 5 as well. The behavior of the acquired spectrum can be reproduced exactly by considering a Hermite-Gauss profile with transverse electromagnetic modes TEM01, TEM01, and TEM11, and by using the parameters $R_M$ = 99%, $R_{WGP}$ = 97.9%. This alteration in the WGP characteristic value is due to the 70° at which we are operating.

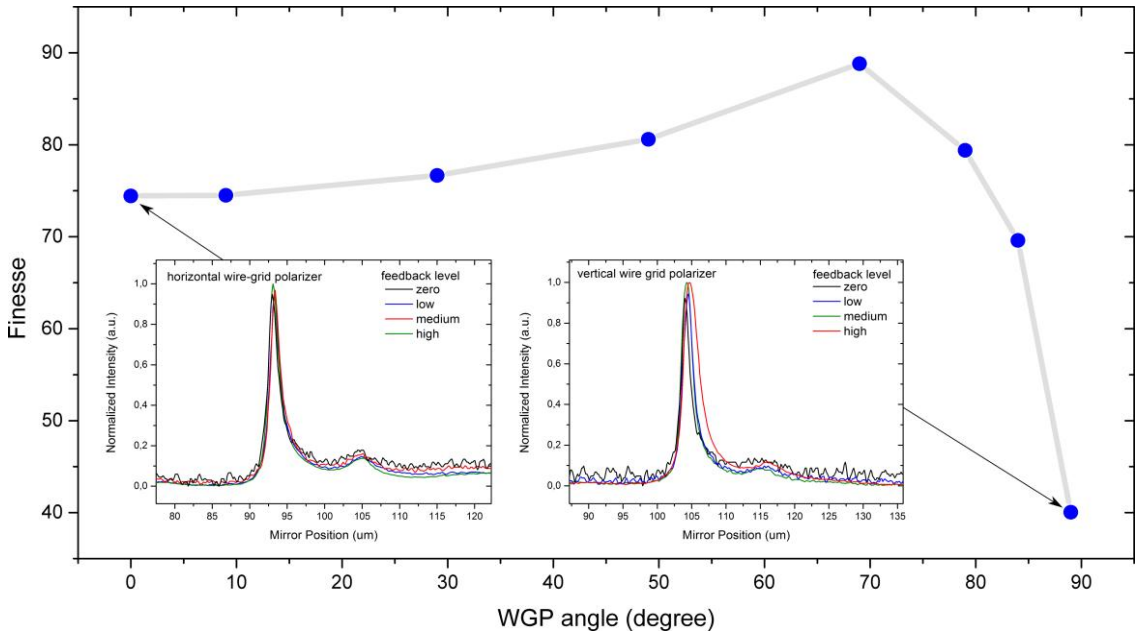

Figure 4. Systematic analysis of the measured BTC finesse for different polarization orientations inside the cavity, i.e., different orientation of the WGP wires. The reported finesse values are still non-perfectly optimized. However, the two effects of optical feedback (OF) and additional cavity losses are visible. The insets show in detail the analysis of the peak width for different beam attenuations, corresponding to different feedback levels, when the wires of the WGP are vertically (inset-right) and horizontally (inset-left) oriented.

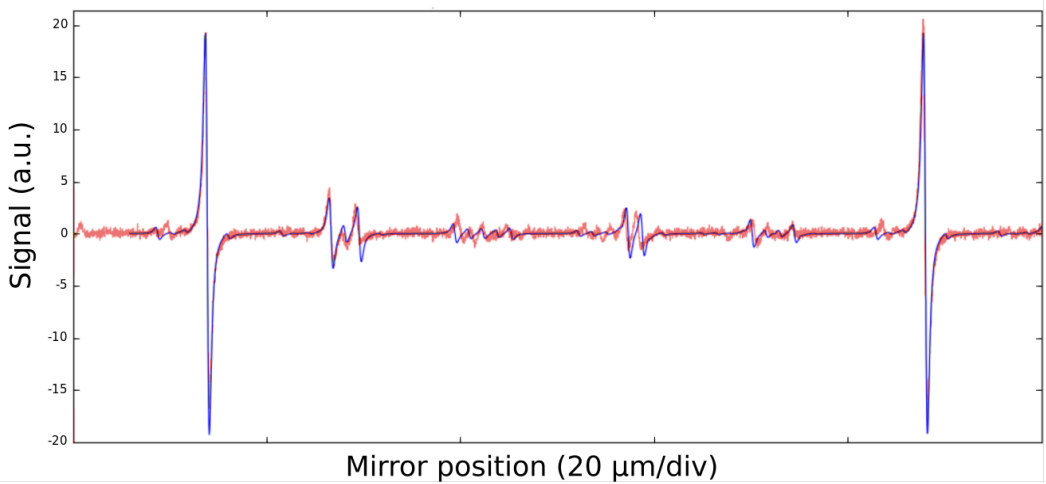

Figure 5. Comparison between experimental (yellow) and simulated spectrum (blue) of the BTC in the experimental condition ($R_M$ = 99%, $R_{WGP}$ = 97.9%), achieving a finesse value of 123. The simulation is performed with the Finesse [24] software, and by using the parameters $R_M$ = 99%, $R_{WGP}$ = 97.9%.

## 5. Conclusions

In conclusion, in this work a new geometry of THz resonant cavity, i.e., a bow-tie cavity, is presented and tested by coupling it to the radiation of a 2.55 THz QCL. The resonator is based on gold mirrors and a free-standing WGP acting as input/output coupler. After a theoretical discussion of the properties of the cavity, a complete characterization is reported, and the experimentally retrieved parameters are compared with the calculated ones. A measured finesse value of F = 123 proves that this is the first resonator with $Q$ = 5.1·10 [5] at THz frequencies, doubling the result previously achieved with different cavity geometries. The discrepancy between the measured and the expected finesse value is to be attributed to the non-perfectly vertical direction of the properties of the cavity, a complete characterization is reported, and the experimentally retrieved parameters are compared with the calculated ones. A measured finesse value of $F = 123$ proves that this is the first resonator with $Q = 5.1 \cdot 10^5$ at THz frequencies, doubling the result previously achieved with different cavity geometries. The discrepancy between the measured and the expected finesse value is to be attributed to the non-perfectly vertical direction of the WGP wires, that, due to OF on the laser, represents the main limitation to the measurement. From the theoretical predictions the presented cavity setup would be able to reach a finesse value of 166, while the limitation to the measurement can be overcome by trying a different input/output coupler, e.g., thin film plastic beam splitters with a thin gold layer on top. This should in principle return a reflection around 99%, and should avoid feedback effects on the laser, resulting in a measured finesse value of more than 158.

**Author Contributions:** Experimental setup realization, L.C., A.C. and S.B.; conceptualization, L.C., D.M., P.D.N. and S.B.; software, A.C. and S.B.; validation, L.C., D.M. and S.B.; data curation, L.C. and S.B.; QCL fabrication supervision, M.S.V.; writing—original draft preparation, L.C. and A.C.; writing—review and editing, all authors; supervision, P.D.N. and S.B.

**Funding:** The authors acknowledge financial support by: European Union FET-Open grant 665,158—ULTRAQCL project; European Union ERC Consolidator grant 681,379–SPRINT; European ESFRI Roadmap "Extreme Light Infrastructure"–ELI project; European Commission–H2020 Laserlab-Europe, EC grant agreement number: 654148.

**Conflicts of Interest:** The authors declare no conflict of interest.